\documentclass[useAMS,usenatbib,twocolumn]{mn2e}

\usepackage{textcomp,graphicx,amsmath,pdflscape,amssymb,setspace,subfig,rotating,multirow,psfrag,color}
\usepackage[table]{xcolor}

\newcommand\aj{{AJ}}%
\newcommand\araa{{ARA\&A}}%
\newcommand\apj{{ApJ}}%
\newcommand\apjs{{ApJS}}%
\newcommand\aap{{A\&A}}%
\newcommand\aapr{{A\&A~Rev.}}%
\newcommand\mnras{{MNRAS}}%
\newcommand\nat{{Nature}}%
\newcommand\pasa{{PASA}}%
\newcommand{\todo}{\ifmmode {\Huge \bullet} \else {\Huge$\bullet$}\fi}
\newcommand{\E        }[1]{\ifmmode 10^{#1} \else $10^{#1}$\fi}
\newcommand{\til}{\ifmmode \sim \else $\sim$\fi}
\renewcommand{\~} {\ifmmode \sim \else $\sim$\fi}

\newcommand\ltsim{\raisebox{-.5ex}{$\;\stackrel{<}{\sim}\;$}}
\newcommand\gtsim{\raisebox{-.5ex}{$\;\stackrel{>}{\sim}\;$}}

\def\Msun{\ifmmode M_{\odot} \else $M_{\odot}$\fi}
\def\msun{\ifmmode M_{\odot} \else $M_{\odot}$\fi}
\def\Lsun{\ifmmode L_{\odot} \else $L_{\odot}$\fi}
\def\mpyr{\ifmmode \Msun\,{\rm yr}^{-1} \else $\Msun\,{\rm yr}^{-1}$ \fi}

\def\hst{\textit{HST}}

\def\HeII{He\,{\small II}}
\def\HeIII{He\,{\small III}}
\def\HeIIln{\HeII\ $\lambda4686$~\AA{} }

\def\ptwosf{$0.7\times10^{-17}$}
\def\pthreesf{$1.0\times10^{-17}$}
\def\ptwosL{$3.4\times10^{34}$}
\def\pthreesL{$5.1\times10^{34}$}

\title[\HeII\ limits at the site of SN2011fe]{Progenitor constraints on the Type-Ia supernova SN2011fe from pre-explosion {\it Hubble Space Telescope} He\,{\LARGE II} narrow-band observations}

\author[Graur, Maoz, \& Shara]
{Or~Graur,$^{1,2,3}$\thanks{E-mail: orgraur@jhu.edu}
Dan~Maoz,$^4$
and Michael M. Shara$^2$
\\
$^1$Department of Physics and Astronomy, The Johns Hopkins University, Baltimore, MD 21218, USA \\
$^2$Department of Astrophysics, American Museum of Natural History, Central Park West and 79th Street, New York, NY 10024-5192, USA \\
$^3$CCPP, New York University, 4 Washington Place, New York, NY 10003, USA \\
$^4$School of Physics and Astronomy, Tel-Aviv University, Tel-Aviv 69978, Israel \\
}

\begin{document}

\maketitle

%========================================= Abstract =========================================&

\setstretch{1}

\begin{abstract}
\noindent We present {\it Hubble Space Telescope} (\hst) imaging observations of the site of the Type-Ia supernova SN2011fe in the nearby galaxy M101, obtained about one year prior to the event, in a narrow band centred on the \HeIIln emission line. In a `single-degenerate' progenitor scenario, the hard photon flux from an accreting white dwarf (WD), burning hydrogen on its surface over $\sim 1$~Myr should, in principle, create a \HeIII\ Str\"{o}mgren sphere or shell surrounding the WD. Depending on the WD luminosity, the interstellar density, and the velocity of an outflow from the WD, the \HeIII\ region could appear unresolved, extended, or as a ring, with a range of possible surface brightnesses. We find no trace of \HeIIln line emission in the \hst\ data. Using simulations, we set $2\sigma$ upper limits on the \HeIIln luminosity of $L_{\rm HeII}<$\ptwosL\ erg~s$^{-1}$ for a point source, corresponding to an emission region of radius $r < 1.8$~pc. The upper limit for an extended source is $L_{\rm HeII}<1.7\times10^{35}$~erg~s$^{-1}$, corresponding to an extended region with $r\sim11$ pc. The largest detectable shell, given an interstellar-medium density of 1 cm$^{-3}$, has a radius of $\sim6$ pc. Our results argue against the presence, within the $\sim 10^5$~yr prior to the explosion, of a supersoft X-ray source of luminosity $L_{\rm bol}\gtsim3\times10^{37}$~erg~s$^{-1}$, or of a super-Eddington accreting WD that produces an outflowing wind capable of producing cavities with radii of 2--6 pc.  

\end{abstract}


%================================= Section 1 - Introduction =================================&

\section{Introduction}
\label{sec:intro}

Type-Ia supernovae (SNe Ia) are most likely the result of the thermonuclear combustion of a carbon-oxygen white dwarf (WD), but the progenitor systems and the processes that lead to ignition and explosion 
have not been identified (see \citealt*{Maoz2013ARAA} for a review). In the double-degenerate (DD) scenario \citep{Iben1984,Webbink1984}, the progenitor system consists of two WDs that merge due to loss of energy and angular momentum to gravitational waves. In the single-degenerate (SD) scenario \citep{Whelan1973}, a WD grows in mass through stable accretion from a non-degenerate companion star, which can be on the main sequence (MS), a sub-giant, a red giant (RG), or a stripped `He star.'

In the SD scenario, the accretion rate of matter onto the WD can fall into three regimes. When accretion rates are below $\sim 3\times10^{-7}~{\rm M_{\odot}~yr^{-1}}$, a thin degenerate hydrogen layer accumulates on the surface of the WD until it ignites explosively, resulting in a nova eruption. When the accretion rate is only slightly below this limit, intervals between eruptions are of order decades, producing objects known as recurrent novae. It is still unclear whether successive episodes of accretion and eruption lead to a net gain or net loss in WD mass \citep{2001ApJ...558..323H,2011A&A...530A..63P,2013AAS...22123306S}. The steady burning regime, in which the WD, of mass $M_{\rm WD}$, burns hydrogen stably on its surface, is confined to the narrow range $3.1\times10^{-7}(M_{\rm WD}/{\rm M_{\odot}} - 0.54) \ltsim \dot{M} \ltsim 6.7\times10^{-7}(M_{\rm WD}/{\rm M_{\odot}} - 0.45)~{\rm M_{\odot}~yr^{-1}}$ \citep{2007ApJ...663.1269N}. Associated with WDs accreting in this range are the objects known as persistent supersoft X-ray sources, which have typical bolometric luminosities in the range $L_{\rm bol}\sim10^{36}$--$10^{38}$ erg s$^{-1}$ and temperatures of $\sim2$--$9\times10^5$ K \citep{1992A&A...262...97V,1997ARA&A..35...69K}. However, recent hydrodynamical models of this accretion-rate regime have obtained some conflicting results regarding its nature -- steady burning or numerous cycles of nova-like eruptions, and on whether there is a net gain or loss of mass (\citealt*{2013MNRAS.433.2884I,2013arXiv1303.3642N}; \citealt{2013ApJ...777..136W}; Hillman et al., in preparation; see Section~\ref{sec:discuss}, below).

The fate of the SD system in the case of accretion rates above the steady burning limit, which are essentially super-Eddington, is also uncertain. The WD could expand into a RG-like configuration, engulfing the companion and effectively stopping the accretion \citep*{1998ApJ...496..376C}. Alternatively, \citet*{1999ApJ...522..487H} have proposed that the excess mass inflow could be re-directed into a fast, $\sim 1000$~km~s$^{-1}$, outflowing optically thick wind, which would evacuate a low-density cavity around the WD \citep{2007ApJ...662..472B}. In such `rapidly accreting' WDs (e.g., \citealt{2013ApJ...771...13L}), the WD continues to accrete and grow at the stable-burning rate, with a photospheric temperature of $\gtsim 10^5~{\rm K}$. 

While more highly absorbed in X-rays than supersoft X-ray sources (although the amount of absorption may depend not only on the amount of obscuring material between the observer and the WD, but also on its velocity and location along the line of sight; \citealt{2013A&A...549A..32N}), rapidly accreting WDs are still hot enough to photoionize \HeII\ in the surrounding gas. A Str\"{o}mgren sphere of ionized H and He could form around the progenitor system, whether it is a supersoft X-ray source \citep{1994ApJ...431..237R} or a rapidly accreting WD \citep{2013MNRAS.432.1640W}, producing emission in the \HeII\ recombination line at $\lambda4686$~\AA{}. Since the recombination time for this line is $t_{\rm rec} \sim 10^5~(1~{\rm cm}^{-3} / n_{\rm ISM})~{\rm yr}$, where $n_{\rm ISM}$ is the interstellar medium (ISM) number density, the \HeII\ signature could be present even if, for some reason, the nuclear burning on the WD surface had ceased $\sim 10^3$--$10^5$ yr before the SN Ia explosion. \citet{2014arXiv1401.1344J} have searched for the \HeIIln line in the spectra of elliptical galaxies, from ionization of the neutral gas by the integrated emission from a putative population of rapidly accreting WD systems and found that the strength of the detected \HeIIln line was consistent with originating solely from the background population of post-asymptotic giant-branch stars, limiting the contribution of accreting WDs with photospheric temperatures of 1.5--6$\times10^5$ K to 5--10 per cent of the total SN~Ia rate. Another argument against the supersoft X-ray source progenitor scenario comes from the Balmer-dominated shocks observed in SN~Ia remnants, which require the fraction of neutral hydrogen in the circumstellar material around the WD to be relatively high \citep{2012A&ARv..20...49V}.

SN~2011fe in the nearby (6.4 Mpc; \citealt{2011ApJ...733..124S}) galaxy M101 has been the best-studied normal SN~Ia (see \citealt{2013PASA...30...46C} and \citealt{KasenNugentSN2011fe} for reviews). Pre-explosion images, along with early multiwavelength data, have been used to rule out RG and most He stars as binary companions in this event (\citealt{Li2011fe,Nugent2011,2012ApJ...753...22B,2012ApJ...750..164C,2012ApJ...746...21H,2012ApJ...751..134M}). The non-detection of radio synchrotron emission at stringent upper limits esentially rules out Roche-lobe overflow accretion within 100--1000~yr of the explosion, at any plausible level, given that with even just a 1-per-cent mass `spillover,' interaction of the SN ejecta with this material would have been detected \citep{2012ApJ...750..164C}. However, material from an accretion flow from a MS companion that had ceased earlier than this would not have been detected. Furthermore, the conclusions depend on the assumed wind velocities and fractions of the post-shock energy density in the circumstellar medium that are in relativistic electrons and magnetic fields ($\epsilon_B$). For example, the limits set by \citet{2012ApJ...750..164C} on the density of a uniform ISM at radii of $10^{15}$--$10^{16}$ cm from the WD scale as $\epsilon_B^{-0.9}$, with $n_{\rm ISM}\ltsim6$ cm$^{-3}$ for $\epsilon_B=0.1$.

In this Letter, we report a {\it Hubble Space Telescope} (\hst) non-detection, and upper limits, on the brightness of the \HeIIln line in a pre-explosion image at the site of SN~2011fe, obtained in 2010, about a year before the event. In Section~\ref{sec:obs}, we describe the observations. We measure detection limits on the \HeIIln luminosity at the SN~2011fe position in Section~\ref{sec:limmag}. In Section~\ref{sec:discuss}, we discuss what constraints these limits place on the progenitor system of SN2011fe. 

%================================= Section 2 - Observations =================================&

\section{Observations}
\label{sec:obs}

M101 was observed with the Wide Field Camera 3 (WFC3) narrow-band F469N filter under \hst\ program GO--11635 (PI: M. Shara), on 2010 February 25 and 2010 April 4--5, 7--9, and 11, with the original objective to search for emission-line signatures of Wolf-Rayet stars \citep{2013AJ....146..162S}. The 50~\AA-wide F469N filter is centred at wavelength $\lambda_0\approx4688$~\AA{}, which includes all of the emission from the \HeIIln line anywhere in the disc of M101 (recession velocity 240 km~s$^{-1}$). Each field of M101 was imaged with two F469N orbits per pointing, for a total exposure time of 6106 s per field. The location of SN2011fe ($\alpha=14^h03^m05^s.733$, $\delta=+54^{\circ}16^{\prime}25^{\prime\prime}.18$; J2000) was covered by field M101-Q, which is centred at $\alpha=14^h03^m11^s.936$, $\delta=+54^{\circ}17^{\prime}08^{\prime\prime}.50$. The line-of-sight extinction towards the location of SN2011fe in WFC3 filters close to F469N is $A({\rm F438W})=0.032$ and $A({\rm F475W})=0.029$ mags \citep{2011ApJ...737..103S}, from which we estimate a similarly negligible extinction in F469N (see also fig.~2 of \citealt{Nugent2011}).

\begin{figure}
 \includegraphics[width=0.477\textwidth]{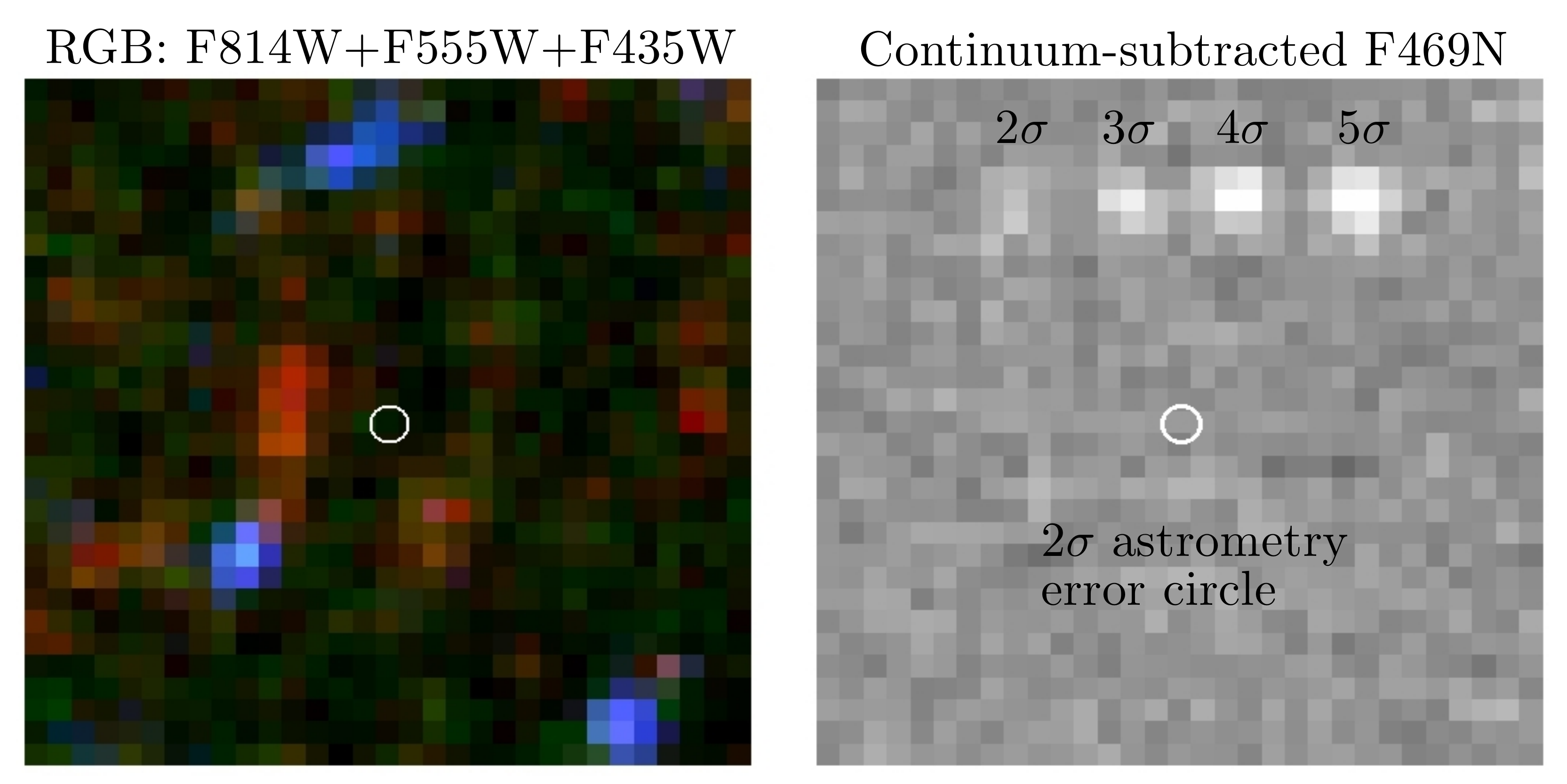}
 \caption{The location of SN~2011fe in (left) an RGB image composed of ACS images in the F814W, F555W, and F435W bands; and (right) in the continuum-subtracted WFC3 F469N image. In the latter image, we
   detect no source within the $2\sigma$ error circle around the location of SN2011fe, as measured by \citet{Li2011fe}, down to $2\sigma$ and $3\sigma$ limiting line fluxes of \ptwosf\ and \pthreesf\ erg s$^{-1}$~cm$^{-2}$, respectively, for a point source. Artificial point sources with S/N ratios of 2--5 are shown for comparison above the error circle. Each of the panels is 1.5 arcsec on a side; North is up and East is left. The scalings of the images that compose the RGB image were chosen to approximate fig.~1 in \citet{Li2011fe}.}
 \label{fig:loc}
\end{figure}

In order to isolate the \HeIIln line from the F469N continuum, \citet{2013AJ....146..162S} scaled and subtracted an image of the same area in the broad-band F435W filter (filter+system central wavelength
$\lambda_0\approx4297$~\AA{}) taken by the \hst\ Advanced Camera for Surveys (ACS) under program GO--9490 (PI: K. Kuntz), which also imaged M101 in the F555W and F814W bands (filter+system central wavelengths
$\lambda_0 \approx 5346$ and $8333$~\AA{}, respectively) on 2002 November 13, and 15--16. Figure~\ref{fig:loc} shows the location of SN2011fe in the continuum-subtracted F469N image, along with an RGB image of the same area, composed of ACS F814W, F555W, and F435W images.

%================================= Section 3 - Detection Limits =================================&

\section{Detection Limits}
\label{sec:limmag}

We find no apparent source of \HeII\ emission at or around the location of SN~2011fe, as shown in Fig.~\ref{fig:loc}. To set constraints on the emission from a progenitor system, we evaluate the fluxes and emission geometries we could expect for various physical progenitor scenarios.

\citet{1994ApJ...431..237R} have calculated photoionization models for the nebulae expected around supersoft X-ray sources from WDs accreting at a rate within the steady-burning range. The bolometric luminosity due to nuclear burning is $L_{\rm bol}=\eta_H X \dot M c^2$, where $\eta_H=0.0069$ is the mass-to-energy conversion efficiency of hydrogen burning, $X=0.72$ is the Solar hydrogen mass abundance, $\dot M$ is the mass accretion rate onto the WD, and $c$ is the speed of light. In this regime, the \HeIII\ is produced in the surrounding ISM, and \citet{1994ApJ...431..237R} show that the \HeIIln line luminosity, for the observed range of supersoft source temperatures of 2--$7\times 10^5$~K \citep{1992A&A...262...97V}, is 
\begin{equation}
L_{\rm He\,II}\approx 1.3\times 10^{-3}~L_{\rm bol} = 2.0 \times 10^{35}~{\rm erg~s}^{-1}(\dot M /\dot M_{\rm max}),
\end{equation}
where $\dot M_{\rm max} = 6.2 \times 10^{-7}~{\rm M_\odot}~{\rm yr}^{-1}$ is the maximum stable-hydrogen-burning accretion rate onto a $1.38~{\rm M_\odot}$ WD. The \HeIII\ region radius (where half of the He is \HeIII~ and half is \HeII) is 
\begin{multline}
r_{\rm str\ddot{o}m}\approx 5~{\rm pc} \left(\frac{L_{\rm bol}}{10^{38}~{\rm erg~s}^{-1}}\right)^{0.35}\left(\frac{n_{\rm ISM}}{10~{\rm cm}^{-3}}\right)^{-0.65} \\
= 6~{\rm pc} \left(\frac{\dot M}{\dot M_{\rm max}}\right)^{0.35}\left(\frac{n_{\rm ISM}}{10~{\rm cm}^{-3}}\right)^{-0.65} ,
\end{multline}
(similar to the $L^{1/3}$ and $n^{-2/3}$ dependences expected, from simple considerations, for a Str\"{o}mgren radius). 

At mass deposition rates higher than $\dot M_{\rm max}$, in the context of the `rapidly accreting WD' scenario, only a rate $\dot M_{\rm max}$ will actually be accreted and burned, with any excess blown off in a fast, $v_w\sim 10^3$~km~s$^{-1}$, wind. The source photospheric temperature is expected to still be $\sim 2\times 10^5~{\rm K}$ \citep{1999ApJ...522..487H}, and there is little absorption of \HeII\ ionizing photons in the outflowing wind, up to mass outflow rates of $\sim 3\times 10^{-6}~{\rm M_\odot}~{\rm yr}^{-1}$ \citep{2013MNRAS.432.1640W}, and hence the \HeIIln luminosity will remain constant at $L_{\rm He\,II}\approx 2.0\times 10^{35}{\rm erg~s}^{-1}$. The region around the WD, however, will be evacuated by the fast wind, producing a very-low-density, $n_{\rm ISM}\sim 10^{-3}$~cm$^{-3}$, wind-blown cavity. The cavity extends out to the radius of the shocked ISM, which is, in turn, surrounded by the unperturbed ISM \citep*{1992ApJ...388...93K,2005ApJ...632..847M}. \citet{2007ApJ...662..472B} have performed numerical hydrodynamical simulations specifically for the case of rapidly accreting WDs, showing how the cavity structures depend on the duration, mass outflow rate, and velocity of the ouflow, as well as on the ISM pressure. For a range of plausible parameters, the bubbles have cavity radii $r_{\rm cav}\sim 10$--40~pc. \citet{2007ApJ...662..472B} find that the expected cavities are incompatible with the observed dynamics and X-ray spectra of most SN~Ia remnants (but see \citealt{2011ApJ...741...96W} for a possible exception). Here, we point out that the WD's ionizing radiation will be absorbed within a spherical shell at the radius where the wind meets the ISM, and hence the \HeIIln line emission will come from this region. 

We have simulated the expected appearance of \HeIIln emission in M101 in the two scenarios -- supersoft sources accreting at $\dot M \le \dot M_{\rm max}$, resulting in \HeIII\ Str\"{o}mgren spheres; 
and rapidly accreting WDs with $\dot M > \dot M_{\rm max}$, that carve out wind-blown bubbles bordered by photoionized \HeIII\ shells. To set limits on \HeIII\ Str\"{o}mgren spheres from a supersoft progenitor, we have planted in the \hst\ image, at the location of SN~2011fe, artificial sources with Gaussian radial profiles, and with half-width-at-half-maximum (HWHM) values ranging from unresolved ($r<1.8$~pc) up to $r=20$~pc, and with fluxes corresponding to a range of luminosities, up to the maximum steady-state-burning \HeII\ luminosity.  
  
We inserted each artificial source in a cutout of the image, centred on the location of SN2011fe, creating, for each value of HWHM, a library of images including sources with different fluxes. We concluded that visual inspection was the best way to search for complex patterns such as rings, and to quantify their detectability. We drew images at random from the various libraries and recorded which sources were detected. Every combination of flux and HWHM was examined ten times, so that we could compute the fraction of times each source was detected. For each HWHM value, this procedure resulted in a data set of the detected fraction of sources as a function of the flux in the source. Each of these data sets was then fit with a cubic spline, resulting in a series of declining detection-efficiency curves. Finally, from each curve, we took the flux at which the detection efficiency dropped to 50 per cent as our detection limit for a source with that HWHM value. For the case of an unresolved point source of line emission, we also performed aperture photometry using a $3\times3$ pixel$^2$ box aperture (which covers $\approx77$ per cent of the point spread function) on hundreds of random, blank locations in the continuum-subtracted F469N image and estimated the mean noise in the image as the root-mean-square (RMS) of the resultant histogram of fluxes. The $2\sigma$ and $3\sigma$ detection limits, defined as the line fluxes at which the signal-to-noise (S/N) ratios of a point source would be 2 and 3, are \ptwosf\ and \pthreesf\ erg s$^{-1}$ cm$^{-2}$. The latter is very similar to the flux at which the detection efficiency in our previous simulation reaches a level of 50 per cent, confirming the bounds from the first simulation as effective $3\sigma$ limits. These detection limits correspond to point-source luminosity limits of $<$\ptwosL\ and $<$\pthreesL\ erg s$^{-1}$, respectively. We have used the RMS of the noise to show, in Figure~\ref{fig:loc}, the expected appearance of point sources with S/N ratios of 2--5 in the \hst\ image.
  
\begin{figure}
 \includegraphics[width=0.477\textwidth]{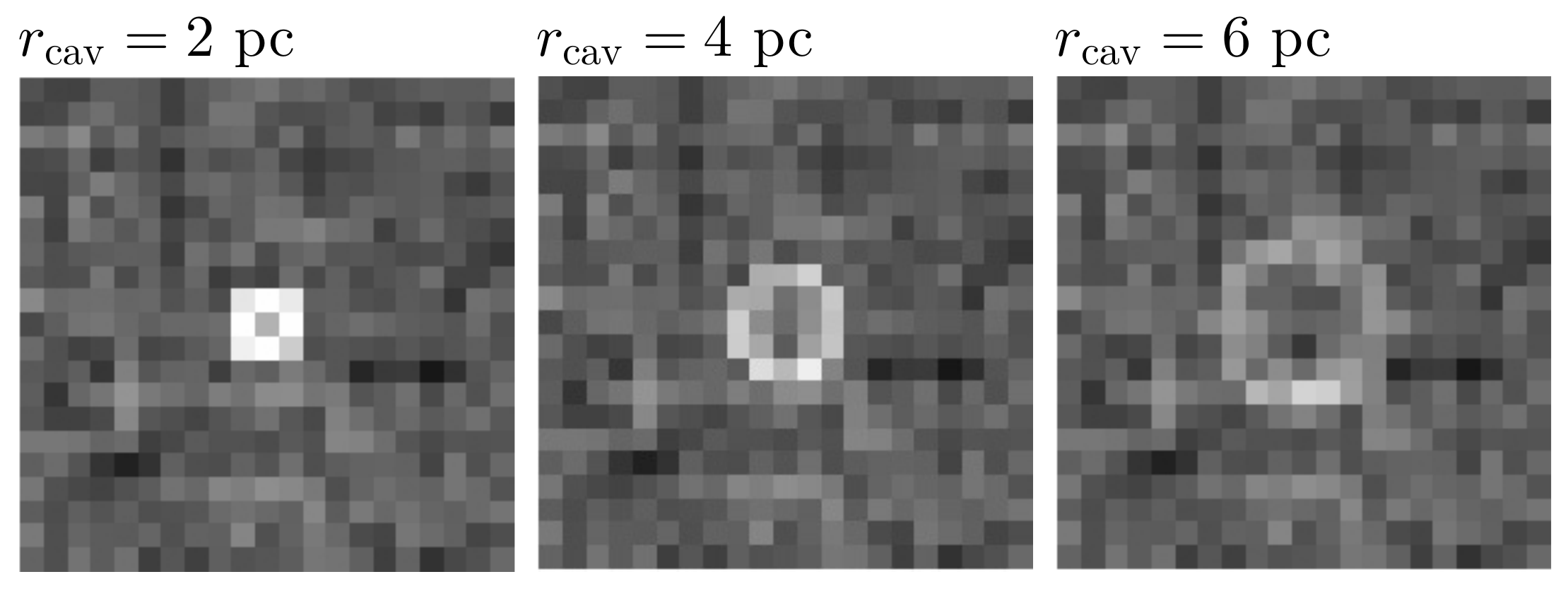}
 \caption{Examples of artificial rings with different radii and \HeIIln line fluxes. From left to right, we show rings with a \HeIIln line flux of $2\times10^{-17}$ erg s$^{-1}$ cm$^{-2}$ (corresponding to a luminosity of $\sim10^{35}$ erg s$^{-1}$) with cavity radii of $r_{\rm cav}=2$, 4, and 6 pc, respectively. All of the rings are centred on the location of SN2011fe in the continuum-subtracted WFC3 F469N image. Each panel is $\sim1$ arcsec on a side; North is up and East is left.}
 \label{fig:ring_eff}
\end{figure}

For the case of a rapidly accreting WD progenitor, we have simulated the appearance of \HeIIln shells, with total luminosities of $2\times 10^{35}$~erg~s$^{-1}$ (corresponding to an effective accretion/burning rate of $\dot M_{\rm max}$), and with inner radii in the range $r_{\rm cav}=2$--40 pc. An optically thin shell of \HeIIln emitting gas will appear, in projection, as an edge-brightened ring, with inner radius $r_{\rm cav}$ and thickness $\Delta r$. The fraction of the luminosity from the shell that is within the projected ring will simply be the corresponding fraction of the volume.
The volume of the shell is
\begin{equation}
 V_{\rm shell} = \frac{4\pi}{3} \left[(r_{\rm cav} + \Delta r)^3 - r_{\rm cav}^3\right].
\end{equation}
From conservation of the number of ionizing photons in the shell, its volume must equal the volume of the Str\"{o}mgren sphere, of radius $r_{\rm str\ddot{o}m}$, for $\dot M = \dot M_{\rm max}$ and a given uniform density, and thus
\begin{equation}
 \Delta r = (r_{\rm str\ddot{o}m}^3 + r_{\rm cav}^3)^{1/3} - r_{\rm cav}.
\end{equation}
The volume within the projected ring is
\begin{multline}
 V_{\rm ring} = 4\pi\int_{r_{\rm cav}}^{r_{\rm cav} + \Delta r} \sqrt{(r_{\rm cav} + \Delta r)^2 - r^2}~ r dr \\
 = \frac{4\pi}{3}[(r_{\rm cav} + \Delta r)^2 - r_{\rm cav}^2]^{3/2}.
\end{multline}
Since, for $\dot M = \dot M_{\rm max}$, $r_{\rm str\ddot{o}m} = 6~{\rm pc}~({n_{\rm ISM}}/{10~{\rm cm}^{-3}})^{-0.65}$, the fraction of the luminosity within the projected ring depends on both $n_{\rm ISM}$ and $r_{\rm cav}$. To determine the detectability of such shells, we have simulated the rings of \HeIIln emission that would appear for a range of values of $r_{\rm cav}$ and $n_{\rm ISM}$. We have again measured the efficiency of detecting these rings as a function of line flux for each ring with a given $r_{\rm cav}$ value and taken the flux at which our detection efficiency reached 50 per cent as our detection limit. Figure~\ref{fig:ring_eff} shows examples of resolved rings with various cavity radii.% and \HeIIln line fluxes. 

\begin{figure}
 \includegraphics[width=0.477\textwidth]{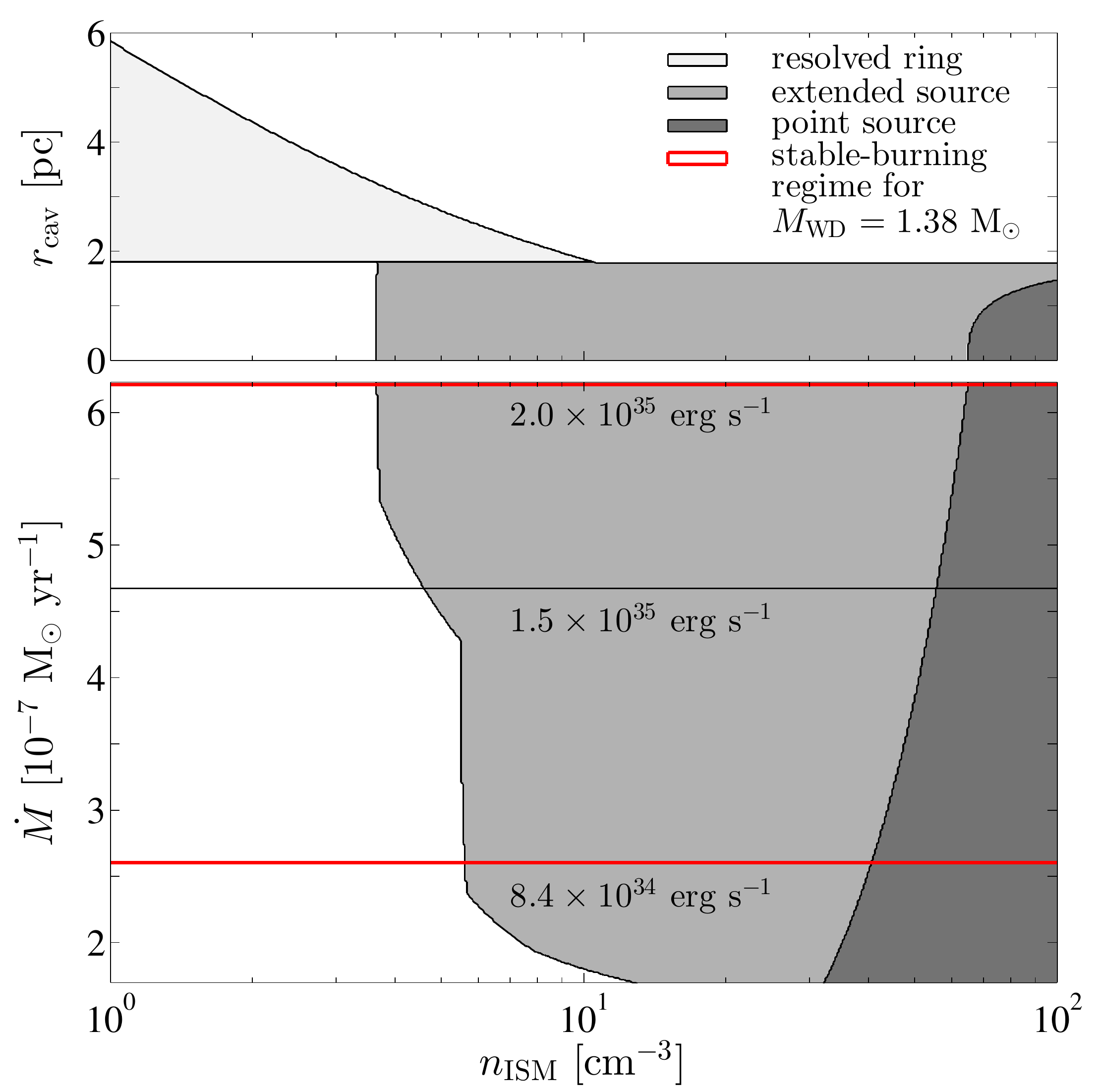}
 \caption{Regions of parameter space ruled out (shaded areas) by the non-detection of \HeIIln emission at the site of SN~2011fe. The bottom panel shows the $\dot M$ vs. $n_{\rm ISM}$ parameter space for the cases of point sources, or extended sources of varying radii. The red curves delineate the lower ($8.4\times10^{34}$ erg s$^{-1}$) and upper ($2.0\times10^{35}$ erg s$^{-1}$) bounds on \HeIIln emission from
 a 1.38 M$_\odot$ WD accreting within the stable-burning regime. The upper panel represents a rapidly-accreting 1.38 M$_\odot$ WD with $\dot M>\dot M_{\rm max}$ and shows the parameter space of $r_{\rm cav}$ vs. $n_{\rm ISM}$, with the observed limits we have set on resolved rings of various radii.}
 \label{fig:nM}
\end{figure}

Figure~\ref{fig:nM} shows our detection limits for both the supersoft X-ray-source and the rapidly-accreting WD scenarios, plotted in the plane of accretion rate, $\dot M$, and ISM density, $n_{\rm ISM}$ (for the first scenario), and in the plane of cavity radius vs. ISM density (for the latter scenario). The shaded areas of this parameter space are excluded by the data. In the supersoft cases (lower panel), low ISM densities lead to large \HeIII\ Str\"{o}mgren spheres, with correspondingly low surface brightness that is difficult to detect in the \hst\ data (lower-left white region in Fig.~\ref{fig:nM}). In the case of rapidly accreting WDs (upper panel), for cavity radii below the \hst\ resolution limit, the detectability  will be essentially like that of the Str\"{o}mgren spheres in the supersoft case (lower-left white 
region in upper panel of Fig.~\ref{fig:nM}). For larger cavities but low ISM densities, $\Delta r/r_{\rm cav}$ is substantial enough so that a large fraction of the shell volume is in the ring, while the ring is still small enough to have detectable surface brightness (triangular grey region in Fig.~\ref{fig:nM}). However, at high densities, $\Delta r$ shrinks (the \HeII\ ionizing photons are absorbed within a geometrically thin shell) and hence the luminosity within the apparent ring becomes too low for detection (upper-right white region of Fig.~\ref{fig:nM}).

Based on Fig.~\ref{fig:nM}, the data rule out a supersoft progenitor system at the site of SN~2011fe within the $t_{\rm rec} \sim 10^5~(1~{\rm cm}^{-3} / n_{\rm ISM})~{\rm yr}$ before the explosion, unless the ISM density within $r\sim 11$~pc of the progenitor is $n<4$--5~cm$^{-3}$. The largest detectable shells have a radius of $\sim 6$~pc at ISM densities of $\sim 1$ cm$^{-3}$. Thus, we can rule out a rapidly accreting WD progenitor for SN~2011fe, as long as the wind-blown cavity it produced has a radius smaller than this.

%================================= Section 4 - Discussion =================================&

\section{Discussion}
\label{sec:discuss}
We have detected no \HeIIln emission at or around the location of SN2011fe. By planting artificial sources in the \hst\ image, simulating resolved and unresolved nebulae, as well as wind-excavated bubbles projected as resolved rings, we have tested both supersoft X-ray and rapidly-accreting WD progenitors. We have measured a $2\sigma$ detection limit of \ptwosf\ erg s$^{-1}$ cm$^{-2}$ for an  unresolved \HeIII\ Str\"{o}mgren sphere (i.e., with a HWHM radius of $<1.8$ pc), corresponding to luminosity limits of $L_{\rm He\,II} <$\ptwosL\ erg s$^{-1}$, or $L_{\rm bol} < 2.6\times10^{37}$ erg s$^{-1}$.  We set an upper limit on the luminosity of a resolved nebula of $L_{\rm He\,II} < 1.7\times10^{35}$ erg s$^{-1}$ (or $L_{\rm bol} < 1.3\times10^{38}$ erg s$^{-1}$), corresponding to a HWHM radius of $\sim11$ pc (i.e., an ISM density of 4 cm$^{-3}$ and $\dot M = \dot M_{\rm max}$). The largest detectable shell has a radius of $\sim 6$ pc, if the ISM density is $\sim1$ cm$^{-3}$. Thus, we rule out a supersoft X-ray source more luminous than $\sim3\times10^{37}$ erg s$^{-1}$ as the progenitor of SN2011fe within the last $10^5$ yr before the SN Ia event, as long as the ISM density is $\gtsim5$ cm$^{-3}$. \citet{Li2011fe} and \citet*{2012MNRAS.426.2668N} have set direct upper limits on the X-ray luminosity at the site of SN2011fe, using pre-explosion {\it Chandra} data from the decade before the event, that are lower than ours by an order of magnitude. However, as noted, our limits apply to a much longer period before the explosion, during which accretion may have ceased.

While our results, above, place limits on specific SN~Ia progenitor scenarios that have been envisaged, an important caveat is that known supersoft X-ray sources apparently do not display the ionization nebulae that one expects in this picture. \citet*{1995ApJ...439..646R} imaged 10 supersoft X-ray sources in the Magellanic Clouds, but detected emission lines around only one source, Cal 83. For Cal 83, 
with $L_{\rm bol}>3\times10^{37}$ erg s$^{-1}$, \citet{2012A&A...544A..86G} detected asymmetric \HeIIln\ emission with a flux of $\sim30\times10^{-16}$ erg s$^{-1}$ cm$^{-2}$. At a distance of $\sim 55$ kpc \citep{1988MNRAS.233...51S}, this corresponds to a luminosity of $\sim 10^{33}$ erg s$^{-1}$, an order of magnitude less than expected from the \citet{1994ApJ...431..237R} models. \citet{1995ApJ...439..646R} measured also H$\alpha$ and [O\,{\small III}] $\lambda5007$~\AA{} emission line fluxes from the nebula around Cal 83. \citet{2012A&A...544A..86G} detected several Balmer, [O\,{\small I}], [O\,{\small II}], [O\,{\small III}], [N\,{\small II}], and [S\,{\small II}] lines. However, none of the \citet{1994ApJ...431..237R} models provided a good fit to the measured line fluxes.

It is unclear why ionization nebulae are not seen in nine out of ten supersoft X-ray sources, nor why, in the one case where \HeII\ line emission is detected, it is weak and asymmetric. One possible explanation is that the supersoft source is surrounded by a disk of absorbing material that, when viewed pole-on, would allow us to observe the X-rays emitted by the WD, but would absorb the \HeII-ionizing photons emitted on the plane of the sky \citep{2013A&A...549A..32N}. Alternatively, it is possible that, contrary to the traditional thinking, there is no true `steady-hydrogen-burning' accretion range onto WDs. On the one  hand, some recent hydrodynamical models of near-Chandrasekhar-mass WDs accreting in this mass range have confirmed the steady-burning picture \citep{2013arXiv1303.3642N,2013ApJ...777..136W}, albeit warning that eventual ignition of the helium ash could eject most or all of the accumulated mass. On the other hand, \citet{2013MNRAS.433.2884I} find, rather than steady hydrogen burning, 1--10-year cyclical nova-like eruptions, but with little mass loss. After thousands of these eruptions, however, a helium eruption will eject most of the gained mass. Finally, Hillman et al. (in preparation) also obtain thousands of hydrogen eruptions with month-to-year-long intervals, but with a significant mass loss, yet with a net gain in mass, all the way up to the Chandrasekhar mass and to explosion as a SN~Ia. A quasi-steady mass outflow from such multiple eruptions, effectively a fast wind, perhaps evacuates a large cavity in the ISM around the WD, much like in the case of rapidly accreting WDs, pushing out any photoionized nebulae to large radii and hence to low and undetectable surface brightness. The partial arc of \HeIIln emission seen in Cal~83 could be from a single high-density partial shell of such nova ejecta, expanding within an otherwise rarified region.

Despite these puzzles, the youth and nearness of SN~2011fe make it worthwhile to search for additional emission line signatures of the progenitor system, even now, after the explosion. At a velocity of $\sim10^4$ km s$^{-1}$, the ejecta of SN2011fe have expanded, to date, to a radius of no more than $\sim 0.015$~pc. Ionizing photons from the SN have reached a radius, in the plane of the sky, of $\sim 0.5$~pc. Thus, a Str\"{o}mgren sphere or shell with \HeIIln luminosity below our detection limits would still be unperturbed by the remnant, as long as the outer radius of the nebula, or the cavity radius of the shell, extended beyond 0.5 pc. While the \HeIIln line is expected to have only 0.13 per cent of the bolometric luminosity of the supersoft source, other lines are expected to be much brighter, e.g. the [O\,{\small III}] $\lambda\lambda4960,5007$~\AA{} doublet and the [N\,{\small II}] $\lambda6585$~\AA{} lines should have $\sim4$ and $\sim1.7$ per cent of $L_{\rm bol}$, respectively. It would be instructive to search for these emission lines in narrow-band images of SN~2011fe, as we have done here for the \HeIIln line. Although M101 has been imaged with \hst\ in the corresponding narrow-band filters, none of these observations have covered the location of SN~2011fe.

%================================= Acknowledgments =================================&

\section*{Acknowledgments}
We thank Carles Badenes, Joanne Bibby, Jennifer Sokoloski, Tyrone Woods, David Zurek, and the anonymous referee for their helpful comments and discussions.
DM acknowledges supported by a grant from the Israel Science Foundation and from ICORE center of excellence of the ISF/PDB.
This research has made use of the NASA/IPAC Extragalactic Database (NED) which is operated by the Jet Propulsion Laboratory, California Institute of Technology, under contract with the National Aeronautics and Space Administration.

%================================= References =================================&

\end{document}